\documentclass[12pt,preprint]{aastex}
\usepackage{ulem}
\usepackage{color}
\usepackage{amsmath}

\begin{document}

\title{Search for electron antineutrinos associated with gravitational wave events GW150914 and GW151226 using KamLAND}

\author{
%
A.~Gando\altaffilmark{1}, 
Y.~Gando\altaffilmark{1}, 
T.~Hachiya\altaffilmark{1}, 
A.~Hayashi\altaffilmark{1},  
S.~Hayashida\altaffilmark{1},
H.~Ikeda\altaffilmark{1},
K.~Inoue\altaffilmark{1,2},
K.~Ishidoshiro\altaffilmark{1},
Y.~Karino\altaffilmark{1},
M.~Koga\altaffilmark{1,2},
S.~Matsuda\altaffilmark{1},
T.~Mitsui\altaffilmark{1},
K.~Nakamura\altaffilmark{1,2}, 
S.~Obara\altaffilmark{1}, 
T.~Oura\altaffilmark{1}, 
H.~Ozaki\altaffilmark{1}, 
I.~Shimizu\altaffilmark{1}, 
Y.~Shirahata\altaffilmark{1}, 
J.~Shirai\altaffilmark{1},
A.~Suzuki\altaffilmark{1}, 
T.~Takai\altaffilmark{1},
K.~Tamae\altaffilmark{1},
Y.~Teraoka\altaffilmark{1},
K.~Ueshima\altaffilmark{1},
H.~Watanabe\altaffilmark{1},
A.~Kozlov\altaffilmark{2},
Y.~Takemoto\altaffilmark{2,3},
S.~Yoshida\altaffilmark{3},
K.~Fushimi\altaffilmark{4},
%
%
%
A.~Piepke\altaffilmark{5,2},
T.~I.~Banks\altaffilmark{6,7},
B.~E.~Berger\altaffilmark{7,2},
B.K.~Fujikawa\altaffilmark{7,2},
T.~O'Donnell\altaffilmark{6,7},
J.G.~Learned\altaffilmark{8},
J.~Maricic\altaffilmark{8}, 
M.~Sakai\altaffilmark{8},
L.~A.~Winslow\altaffilmark{9},
E.~Krupczak\altaffilmark{9},
J.~Ouellet\altaffilmark{9},
Y.~Efremenko\altaffilmark{10,11,2},
H.~J.~Karwowski\altaffilmark{12,13},
D.~M.~Markoff\altaffilmark{12,14},
W.~Tornow\altaffilmark{12,15,2},
J.~A.~Detwiler\altaffilmark{16,2},
S.~Enomoto\altaffilmark{16,2},
M.P.~Decowski\altaffilmark{17,2}
}

\affil{The KamLAND Collaboration}

\altaffiltext{1}{Research Center for Neutrino Science, Tohoku University, Sendai 980-8578, Japan}
\altaffiltext{2}{Kavli Institute for the Physics and Mathematics of the Universe (WPI), The University of Tokyo Institutes for Advanced Study, The University of Tokyo, Kashiwa, Chiba 277-8583, Japan}
\altaffiltext{3}{Graduate School of Science, Osaka University, Toyonaka, Osaka 560-0043, Japan}
\altaffiltext{4}{Faculty of Integrated Arts and Science, University of Tokushima, Tokushima, 770-8502, Japan}
\altaffiltext{5}{Department of Physics and Astronomy, University of Alabama, Tuscaloosa, Alabama 35487, USA}
\altaffiltext{6}{Physics Department, University of California, Berkeley, California 94720, USA} 
\altaffiltext{7}{Lawrence Berkeley National Laboratory, Berkeley, California 94720, USA}
\altaffiltext{8}{Department of Physics and Astronomy, University of Hawaii at Manoa, Honolulu, Hawaii 96822, USA}
\altaffiltext{9}{Massachusetts Institute of Technology, Cambridge, Massachusetts 02139, USA}
\altaffiltext{10}{Department of Physics and Astronomy, University of Tennessee, Knoxville, Tennessee 37996, USA}
\altaffiltext{11}{National Research Nuclear University, Moscow, Russia}
\altaffiltext{12}{Triangle Universities Nuclear Laboratory, Durham, North Carolina 27708, USA}
\altaffiltext{13}{The University of North Carolina at Chapel Hill, Chapel Hill, North Carolina 27599, USA}
\altaffiltext{14}{North Carolina Central University, Durham, North Carolina 27701, USA}
\altaffiltext{15}{Physics Department at Duke University, Durham, North Carolina 27705, USA}
\altaffiltext{16}{Center for Experimental Nuclear Physics and Astrophysics, University of Washington, Seattle, Washington 98195, USA}
\altaffiltext{17}{Nikhef and the University of Amsterdam, Science Park, Amsterdam, the Netherlands}

\vspace{+0.2in}

\begin{abstract}
We present a search for low energy antineutrino events coincident with the gravitational wave events GW150914 and GW151226, and the candidate event LVT151012 using KamLAND, a kiloton-scale antineutrino detector. 
We find no inverse beta-decay neutrino events within $\pm 500$ seconds of either gravitational wave signal. This non-detection is used to constrain the electron antineutrino fluence and the total integrated luminosity of the astrophysical sources.

\end{abstract}

\keywords{gravitational waves --- neutrinos}

\maketitle

\section{Introduction}
With the detection of gravitational waves (GW) by the Advanced Laser Interferometer Gravitational-wave Observatory (LIGO)\citep{abbott2016} and high-energy astrophysical neutrinos by IceCube \citep{icecube2015}, the era of multi-messenger astronomy has started in earnest. The combination of these signals with electromagnetic observations offers an unprecedented glimpse into the dynamics of astrophysical phenomena and is already leading to unexpected results.

The first gravitational wave event was observed by LIGO on 14 Sep 2015 at 09:50:45 UTC. Denoted GW150914, this event was observed to have a false alarm rate of less than 1 event per 203,000 years, corresponding to a significance of $>5.1\sigma$ \citep{abbott2016}. 
This likely originated from the coalescence of two black holes at a luminosity distance of $410_{-180}^{+160}$ Mpc \citep{abbott2016}.
The second GW event, GW151226, was observed by LIGO on 26 Dec 2015 at 03:38:53 UTC \citep{abbott2016d}. GW151226 likely originated from a black hole-black hole (BH-BH) merger which took place at a luminosity distance of $440^{+180}_{-190}$ Mpc \citep{abbott2016d}. 

We also analyze a GW candidate, dubbed LVT151012 (LIGO-Virgo-Trigger), which occurred on 12 Oct 2015 at 09:54:43 UTC. While LVT151012 did not cross the threshold required to claim a detection it is unlikely to be a background event, being the only other event reported by LIGO at time of writing to have a $>50\%$ chance of astrophysical origin \citep{LVT2016}. The BH-BH merger suggested by LVT151012 occurred at a luminosity distance of $1100^{+500}_{-500}$ Mpc \citep{abbott2016c}. 

There is no known mechanism for the production of either neutrinos or electromagnetic waves in a BH-BH merger. mini-balloon
While both gamma-ray bursts and neutrino signals can originate from black holes with rapidly accreting disks, the accretion disk is not expected to be present during a BH-BH merger and therefore neither a neutrino signal nor a gamma-ray burst is predicted \citep{caba2011}. However, the {\it Fermi} telescope observed a coincident gamma-ray burst occurring 0.4 seconds after GW150914 with a false alarm probability of 0.0022 \citep{con2016}. There is large uncertainty in the origin region but it is consistent with that reported by LIGO \citep{con2016}. The statistical treatment has been debated and the event may be consistent with background \citep{greiner}. If this burst truly originates from the same black hole merger as GW150914 it could imply that some accretion disk remained during the merger, thus motivating a multi-messenger analysis including neutrinos of all energies. 

In this paper, we search for correlations between these GW events and electron antineutrinos of a few tens of MeV, and place constraints on the neutrino fluence and luminosity. This work is complementary to the multi-messenger analysis performed by IceCube and ANTARES at higher energies, which did not find any neutrino events correlated with GW150914 with sufficient significance \citep{icecube2016}.

\section{KamLAND} 
KamLAND (Kamioka Liquid scintillator Anti-Neutrino Detector) is optimized to search for $\sim$ MeV neutrinos and antineutrinos. KamLAND is located under 2,700 meter-water-equivalent of vertical rock, below Mt. Ikenoyama in Gifu-prefecture, Japan.
KamLAND consists of an 18-m diameter stainless steel sphere which has 1,325 17-inch and 554 20-inch photomultiplier tubes mounted on its inside surface. The sphere contains a 13-m diameter EVOH/nylon outer balloon surrounded by pure mineral oil.  This outer balloon encloses 1 kton of highly purified liquid scintillator.
Surrounding the stainless steel sphere is a cylindrical 3.2 kton water-Cherenkov detector to provide shielding and allow cosmic-ray muon identification. Additional details of KamLAND are summarized in \citep{suzuki2014}. 
During the period corresponding to the detection of GW150914 and LVT151012, a 3.08-m-diameter transparent nylon inner balloon (``mini-balloon'') containing 13 tons of Xe-loaded liquid scintillator had been placed at the center of the detector \citep{gando2012physC}. At the time that GW151226 was detected, the mini-balloon had been removed and KamLAND was in its normal configuration. 

In this analysis, we will focus on the detection of antineutrinos through the inverse beta-decay (IBD) reaction: $\bar \nu_e + p \rightarrow e^+ + n$. 
This process is characterized by a delayed-coincidence event pair signature.
The prompt event is a combination of the deposition of the kinetic energy of the positron and its subsequent annihilation into gamma rays. This event encodes the energy of the incoming antineutrino.
This annihilation occurs on a very short time scale. Because the angular distribution of the positron emission and the subsequent scintillation light are isotropic, KamLAND has no directional sensitivity.
The delayed event is the emission of a gamma-ray when the neutron captures on carbon or a proton, with a mean neutron capture time of $207.5\pm 2.8 \mbox{ }\mu$s \citep{abe2010}.
The detection of this second gamma-ray completes the delayed coincidence pair. 

KamLAND's main background source depends on the energy region. At energies of a few MeV, reactor neutrinos and geoneutrinos dominate for standard analysis \citep{gando2013}. 
For the purposes of this coincidence search we have used the maximum possible detector volume and removed filters which screen out accidental radiation from the support structure. Thus, the background is dominated by low-energy events caused by accidental coincidences of natural radioactivity. The majority of the remaining low-energy backgrounds are from reactor neutrinos and geoneutrinos; the event rate of reactor neutrinos was about 0.1-0.2 events per day during this period, and the geoneutrino background rate was about 0.1 event per day.
Above {$\sim 7.5$ MeV}, the majority of the background comes from neutral current interactions with atmospheric neutrinos \citep{gando2012anu}. 
  This background is effectively constant up to $\sim$100 MeV.  
For a more detailed discussion of KamLAND's backgrounds, please see \cite{Asakura2015} and \cite{gando2013}. 
KamLAND's lower energy threshold gives it an advantage in this low energy neutrino range compared to the IceCube and Super-Kamiokande detectors.

\section{Coincidence Search}
The detection of the gamma-ray burst by the {\it Fermi} telescope suggests that the BH-BH merger detected by LIGO might have retained its accretion disk. 
The spectrum of accretion disk neutrinos around a single black hole is expected to peak around 10 MeV, with the majority falling well below 100 MeV (\cite{caba2011} and \cite{mcla2007}). In the absence of a mechanism for neutrino production by a BH-BH merger, we consider the single BH accretion model and search for IBD events with visible energies between 0.9 and 100 MeV, corresponding to neutrino energies between 1.8 and 111 MeV. 
We choose a coincidence window of $\pm 500$ seconds by selecting the largest expected time gap between GW events and high energy neutrino events described in \citep{baret2011}. 
We use the standard KamLAND event selection \citep{abe2010}. This restricts the analysis to $R<6$~m to remove backgrounds from the main balloon. It also applies a veto of 2~s within a 3~m cylinder or a 2~s full detector veto following muon events depending on the quality of the muon track reconstruction to reduce background due to the long-lived muon spallation product $^{9}$Li/$^{8}$He \citep{abe2010}. The muon veto leads to a difference in the livetime to real time ratio, where livetime is defined as the period of time during which the detector 
was sensitive to $\bar\nu_e$ and includes corrections for calibration periods, detector maintenance, 
and other factors. The average livetime to real time ratio is $\epsilon_\text{live} = 0.89$ for the two 
KamLAND runs containing GW150914 and LVT151012. 
The livetime to real time ratio is $\epsilon_\text{live} = 0.81$ for the run containing GW151226. 

Two changes were made to the standard KamLAND criteria \citep{gando2013};
we removed the mini-balloon cut and the likelihood selection.  These cuts increase
the background due to the mini-balloon and main balloon. The total detection 
efficiency of $\epsilon_\text{s} = 0.93$ was then estimated from Monte Carlo simulation.

\begin{figure}
\centering
\includegraphics[width=9cm]{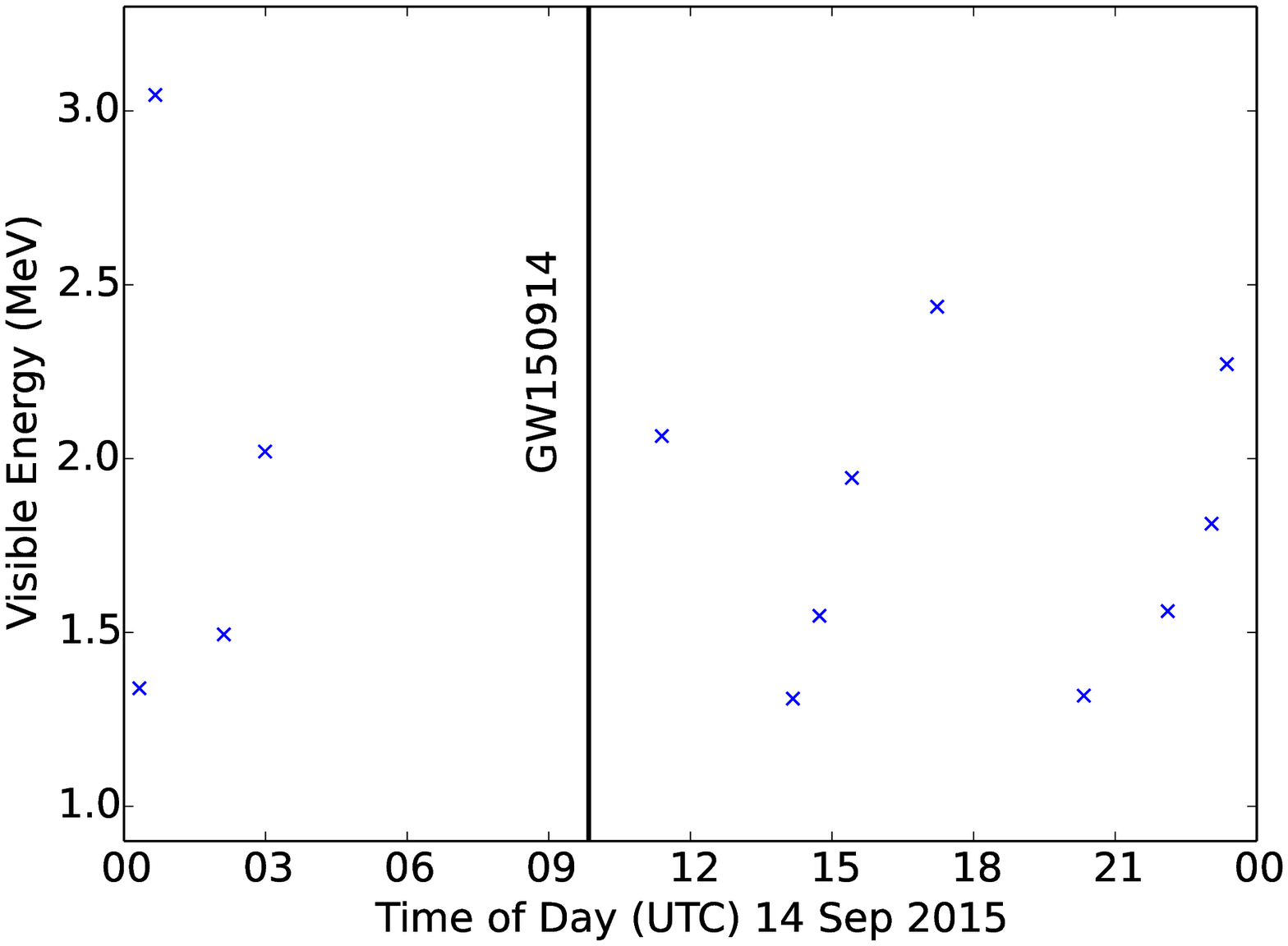}
\caption{Neutrino events between 0.9-100 MeV visible energy occurring on 14 Sep 2015. The highest energy event in this time span was at 3.05 MeV. The time of GW150914 is marked. There were no events within 500 seconds of GW150914. }
\label{GWTime}
\end{figure}

\begin{figure}
\centering
\includegraphics[width=9cm]{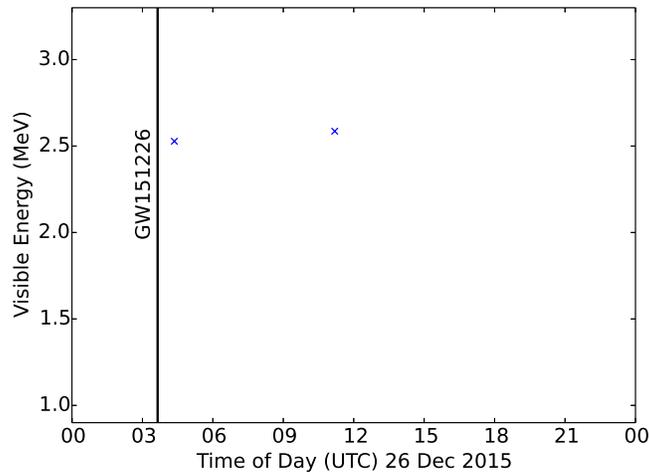}
\caption{Neutrino events between 0.9-100 MeV visible energy occurring on 26 Dec 2015. The highest energy event in this time span was at 2.58 MeV. The time of GW151226 is marked. There were no events within 500 seconds of GW151226. The closest event occurred approximately 40 minutes after GW151226 and was consistent with background.}
\label{GW2Time}
\end{figure}

\begin{figure}
\centering
\includegraphics[width=9cm]{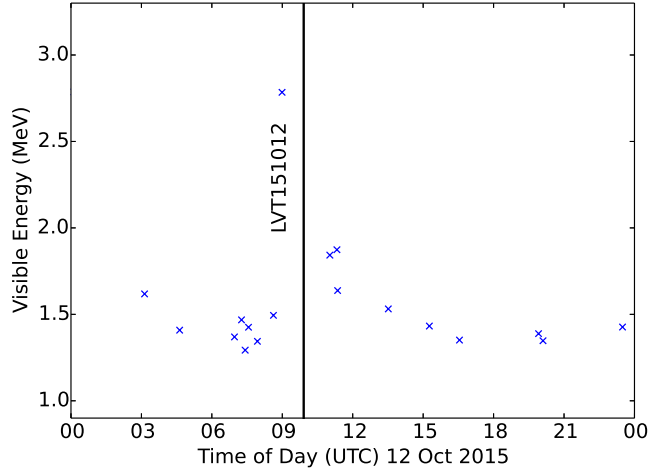}
\caption{Neutrino events between 0.9-100 MeV visible energy occurring on 12 Oct 2015. The highest energy event in this time span was at 2.78 MeV. The time of LVT151012 is marked. There were no events within 500 seconds of LVT151012. The closest event occurred at 1124 seconds prior to LVT151012 and was consistent with background.}
\label{LVTTime}
\end{figure}

We searched for events which fell within a 500 second window of the two GW events GW150914 and GW151226, and the candidate event LVT151012. No events were found within the target window of GW150914 (Figure \ref{GWTime}), GW151226 (Figure \ref{GW2Time}), or LVT151012 (Figure \ref{LVTTime}). 
The closest neutrino candidate event to either GW event occurred 1124 seconds prior to LVT151012. This event was at a very low energy of 1.4 MeV and occurred near the nylon corrugated pipe which supports the mini-balloon. Therefore, this event was consistent with expected background and may have been caused by contamination from the mini-balloon support structure. There were no other events within three hours of LVT151012. 
The closest neutrino candidate to GW151226 occurred about 40 minutes away and was of less than 3 MeV. 
The closest two neutrino candidate events to GW150915 occurred about 2.5 hours away from the event and were both of less than 2 MeV. Therefore all adjacent observed events are likely background. 

The background rate for GW150914 and LVT151012 is given by the average number of IBD events under 100 MeV occurring per second of detector livetime between April 2015 and early November 2015. 
KamLAND's background during this period is $(2.02\pm 0.04) \times 10^{-4}$ events per second of livetime. This corresponds to 0.18 events in a 1000 second real time window.  The accidental coincidence rate during this period was $1.7\times10^{-4}$ events/sec; thus accidental coincidences dominate the background.  

We checked the stationarity of the background rate and found that the event rate was statistically constant.  Using the previously calculated background rate and the detection of no coincidence events, we determined the $90\%$ confidence limit on the number of detected neutrinos is calculated from the background rate to be $N_{90} = 2.26$ using the Feldman Cousins method \citep{feldman}. 

The background rate for GW151226 is given by the average number of IBD events under 100 MeV occurring per second of detector livetime between 23 Dec 2015 and 4 Jan 2016. This time period was chosen to avoid the period during which KamLAND underwent some refurbishment work. The background rate during this period was found to be $(3.4\pm0.6)\times10^{-5}$ events per second of livetime, giving $N_{90} = 2.41$ \citep{feldman}. This background corresponds to 0.03 events per 1000 seconds of real time.  We expect this background to be lower than that for the period from April to November because the mini-balloon, a source of background contamination, was not in the detector during this time period. The accidental coincidence rate during this period was $2.6\times10^{-5}$ events/sec.

\section{Fluence and Luminosity}
We translate our Feldman Cousins upper limit into a fluence upper limit at the detector. 
This fluence upper limit is given in neutrinos per cm$^2$ by
\begin{equation}
\label{fluenceEq}
F_\text{UL}  = \frac{N_{90}}{N_T   \epsilon_\text{live} \epsilon_\text{s}  \int \sigma(E_\nu)\lambda(E_\nu) dE_\nu},
\end{equation}
where 
$N_T$ is the total number of target protons in the fiducial volume, 
$\epsilon_\text{live}$ is the mean livetime to real time ratio, 
$\epsilon_\text{s}$ is the total detection efficiency, 
$\sigma(E_\nu)$ is the total neutrino cross section,
and 
$\lambda(E_\nu)$ is the normalized neutrino energy spectrum \citep{fukuda2002}.
The estimated target number for KamLAND is $N_\text{T} = (5.98 \pm 0.13) \times 10^{31}$. 
%
%
%
The neutrino IBD cross section was taken from \cite{strumia2003}.  The quenching effect and the effect of energy resolution were considered and found to be negligible; thus, these effects are not included in equation (\ref{fluenceEq}). 

The electron antineutrino fluence upper limit without oscillation and assuming a monochromatic spectrum is given by
\begin{equation}
\label{fluenceMono}
F_\text{UL}(E_\nu)  = \frac{N_{90}}{N_\text{T} \sigma(E_\nu) \epsilon_\text{live} \epsilon_\text{s}}.
\end{equation}
The resulting upper limit on fluence ranges from about $10^{13}$ cm$^{-2}$ for a neutrino energy of 1.8 MeV to about $10^8$ cm$^{-2}$ for a neutrino energy of 100 MeV.  The monochromatic spectrums for electron antineutrino fluence upper limit are shown in Figure \ref{fluence}. 


In the absence of a BH-BH merger-specific neutrino energy spectrum prediction, we choose the spectrum given by the normalized pinched Fermi-Dirac distribution for zero chemical potential and pinching factor $\eta = 0$:
\begin{equation}
\label{pinchedFD}
\lambda_{FD}(E) = \frac1{T^3 F_2(\eta)}\frac{E^2}{e^{E/T-\eta}+1},
\end{equation}
where the complete Fermi Dirac integral $F_n(\eta)$ is given by
\begin{equation}
\label{completeF}
F_n(\eta) = \int_0^\infty\frac{x^n}{e^{x-\eta}+1}dx.
\end{equation}
The temperature is given by $T = \langle E \rangle /3.15$. We choose $E = 12.7$ MeV from \cite{Caballero2016}; the small change in average energies between accretion disk models had a negligible impact on the result. 

Substituting this spectrum into (\ref{fluenceEq}) and performing the integration between electron antineutrino energies of $E_\text{min} = 1.8$ MeV and $E_\text{max} = 111$ MeV, we get a total integrated electron antineutrino fluence for both GW150914 and LVT151012 of
\begin{equation}
\label{fluence}
F \leq 3.1\times10^{9} \text{cm}^{-2}.
\end{equation} 
The total integrated electron antineutrino fluence for GW151226 is
\begin{equation}
\label{fluence2}
F \leq 3.6\times 10^{10}  \text{ cm$^{-2}$}. 
\end{equation} 


There is a large uncertainty in the distance for all of GW150914, GW151226 and LVT151012, so the total energy upper limit is here displayed as a function of the true distance to source, $D_\text{GW}$.  
The electron antineutrino total energy upper limits without oscillation for GW150914, GW151226 and LVT151012 are thus given by
\begin{equation}
\label{lumGW}
E_\text{\tiny GW150914} \leq 1.26 \times 10^{60}\left(\frac{D_\text{gw}}{410 \text{ Mpc}}\right)^2 \text{ ergs} 
\end{equation}
and
\begin{equation}
\label{lumGW2}
E_\text{\tiny GW151226} \leq 1.71\times10^{60} \left(\frac{D_\text{gw}}{440 \text{ Mpc}}\right)^2 \text{ ergs}
\end{equation}
and finally, 
\begin{equation}
\label{lumLVT}
E_\text{\tiny LVT151012} \leq 9.06 \times 10^{61} \left(\frac{D_\text{gw}}{1100 \text{ Mpc}}\right)^2 \text{ ergs}.
\end{equation}
This limit complements the upper limit on total energy found by the IceCube-ANTARES joint analysis since the results are based on a different energy region.  The IceCube-ANTARES upper limit on total radiated energy is
\begin{equation}
\label{icecube}
E^\text{ul}_{\nu,\text{ tot}} \sim 10^{52}\text{ - }10^{54}\left(\frac{D_\text{gw}}{410 \text{ Mpc}}\right)^2 \text{ ergs}
\end{equation}
for neutrinos in the $\gg$ GeV energy range \citep{icecube2016}. 

The neutrino event rate scales as a function of the disk's mass accretion rate For current detector masses (for example, Super-Kamiokande), the number is $\sim10,000$ events at 10 kpc, so accounting for the $1/r^2$ scaling with distance, Super-Kamiokande may see one event from a black hole merger at 1 Mpc \citep{Caballero2016}.

Unfortunately, our results do not constrain any viable accretion disk model. \cite{Caballero2016} predicts approximately 500 events per kiloton at 10 kpc for accretion rates on the order of $5 M_\odot$/s. At 1000 Mpc a 100 gigaton detector would be required.  This detector would be similar to IceCube \citep{icecube2016}, but instrumented more densely to obtain a $\sim10$MeV energy threshold. 


\section{Conclusion} 
No coincident neutrino events were found within 500 seconds of either GW150914, GW151226, or LVT151012. 
We determined a monochromatic fluence upper limit, as well as an upper limit on the source luminosity for each GW event and candidate GW event using the standard source model. This places a bound on the total energy released as low energy neutrinos. 
%
The lack of coincident IBD events detected by KamLAND further supports the conclusion by the Dark Energy Survey Collaboration that GW150914 was not a core-collapse supernova in the Large Magellanic Cloud \citep{annis2016}. 

As Advanced LIGO continues its operation, we can expect many more opportunities to perform multi-messenger searches and look for coincidence neutrinos. 
The more complete understanding of the source dynamics provided by such a search grants us an exciting opportunity to explore black holes, supernova, and other elusive astrophysical phenomena.

\section{Acknowledgements}
We are indebted to the LIGO Scientific Collaboration for their gravitational wave observations. 
KamLAND is supported by MEXT KAKENHI Grant Numbers 26104002, 26104007; the World Premier International Research Center Initiative (WPI Initiative), MEXT, Japan; and under the U.S. Department of Energy (DOE) grants no. DE-FG03-00ER41138, DE-AC02-05CH11231, and DE-FG02-01ER41166, as well as other DOE and NSF grants to individual institutions, and Stichting Fundamenteel Onderzoek der Materie (FOM) in the Netherlands. The Kamioka Mining and Smelting Company has provided services for activities in the mine. We thank the support of NII for SINET4.

\bibliographystyle{apj}

\clearpage

\begin{figure}
\centering
\includegraphics[width=10cm]{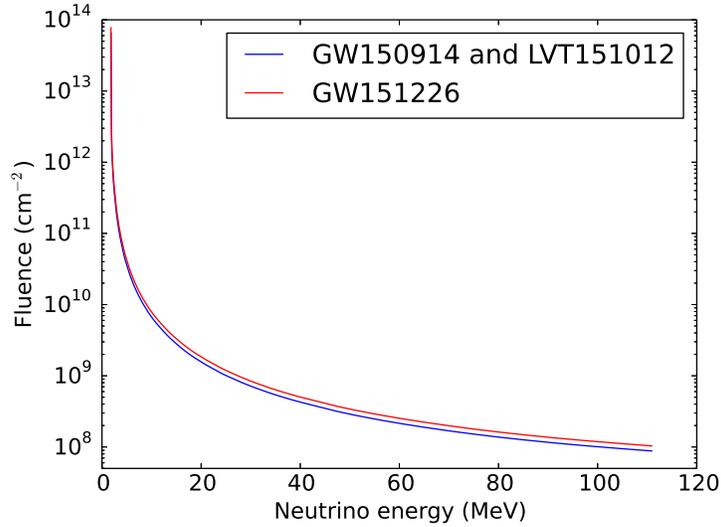}
\caption{Upper limit of electron antineutrino fluence at detector for each energy between 0.9 and 100 MeV assuming a monochromatic spectrum. The spectrums for GW150914 and LVT151012 are the same, due to their shared background rate, and thus they share a fluence upper limit. The background rate for GW151226 is about an order of magnitude lower and thus its fluence upper limit is a bit higher. }
\label{fluence}
\end{figure}

\end{document}